%% file: ScLikePheno.tex
\documentclass[11pt,a4paper]{article}
\usepackage{jheppub}
\pdfoutput=1

\usepackage[utf8]{inputenc}
\usepackage{amssymb}
\usepackage{amsmath}
\usepackage{amsfonts}
\usepackage{graphicx}
\usepackage{xcolor}
\usepackage{xspace}
\usepackage[normalem]{ulem} 
\usepackage{lscape}
\usepackage{ wasysym }
\usepackage{float}
\usepackage{mathrsfs}
\usepackage{slashed}
\usepackage{multirow,rotating}

\usepackage{tikz}

\usepackage{mathrsfs}   
\usepackage{slashed}     
\input{myTikz}

\usepackage{hyperref}
\hypersetup{
  colorlinks=true,
  allcolors=darkpurple,
  pdfborder={0 0 0},
  linktocpage=true
}

\definecolor{darkpurple}{rgb}{0.5,0,0.5}
\definecolor{cambridgeblue}{rgb}{0.64, 0.76, 0.68}
\definecolor{darkraspberry}{rgb}{0.53, 0.15, 0.34}

\def\gsim{\raise0.3ex\hbox{$\;>$\kern-0.75em\raise-1.1ex\hbox{$\sim\;$}}}
\def\lsim{\raise0.3ex\hbox{$\;<$\kern-0.75em\raise-1.1ex\hbox{$\sim\;$}}}

\newcommand{\ba}[1]{\begin{eqnarray} \label{(#1)}}
\newcommand{\ea}{\end{eqnarray}}

\def\gsim{\raise0.3ex\hbox{$\;>$\kern-0.75em\raise-1.1ex\hbox{$\sim\;$}}}
\def\lsim{\raise0.3ex\hbox{$\;<$\kern-0.75em\raise-1.1ex\hbox{$\sim\;$}}}

\makeatletter
\g@addto@macro\bfseries{\boldmath}
\newcommand\Label[1]{&\refstepcounter{equation}(\mathrm{\theequation})\ltx@label{#1}&}
\makeatother

\graphicspath{{figs/}}

\allowdisplaybreaks

\title{Long-lived particle phenomenology in one-loop neutrino mass
  models with dark matter}

\author[a]{Carolina Arbel\'aez,}
\emailAdd{carolina.arbelaez@usm.cl}
\affiliation[a]{{\it Universidad T\'ecnica Federico Santa Mar\'ia
 and Centro Cient\'ifico Tecnol\'ogico\\
de Valpara\'iso CCTVal, Casilla 110-V, Valpara\'iso, Chile}}

\author[b,c]{Giovanna Cottin,}
\emailAdd{gfcottin@uc.cl}
\affiliation[b]{Instituto de F\'isica, Pontificia Universidad Cat\'olica de Chile, \\Avenida Vicu\~{n}a Mackenna 4860, Santiago, Chile}
\affiliation[c]{Millennium Institute for Subatomic Physics at the High Energy Frontier (SAPHIR), Fern\'andez Concha 700, Santiago, Chile}

\author[d,c]{Juan Carlos Helo,}
\emailAdd{jchelo@userena.cl}
\affiliation[d]{Departamento de F\' isica, Facultad de Ciencias, Universidad de La Serena, 
	Avenida Cisternas 1200, La Serena, Chile }

\author[e]{Martin Hirsch,}
\emailAdd{mahirsch@ific.uv.es}
\affiliation[e]{Instituto de F\'{\i}sica Corpuscular
(CSIC-Universitat de Val\`{e}ncia), \\ C/ Catedr\'atico Jos\'e
Beltr\'an 2, E-46980 Paterna (Val\`{e}ncia), Spain}

\author[c,f,g]{T\'essio B. de Melo}
\emailAdd{tessio.melo@uvm.cl}
\affiliation[f]{Universidad Andr\'{e}s Bello, Facultad de Ciencias Exactas, \\ Departamento de Ciencias F\'{i}sicas-Center for Theoretical and Experimental Particle Physics, Fern\'{a}ndez Concha 700, Santiago, Chile}
\affiliation[g]{Universidad Vi\~{n}a del Mar, Escuela de Ciencias, Agua Santa 7055, Rodelillo, Vi\~{n}a del Mar, Chile}

\abstract{

  Neutrino masses and dark matter (DM) might have a common origin. The
  {\em scotogenic} model can be considered the proto-type model
  realizing this idea, but many other variants exist. In this paper we
  explore the phenomemology of a particular DM neutrino mass model,
  containing a triplet scalar. We calculate the relic density and
  check for constraints from direct detection experiments.  The
  parameter space of the model, allowed by these constraints, contains
  typically a long-lived or quasi-stable doubly charged scalar, that
  can be searched for at the LHC. We reinterpret existing searches
  to derive limits on the masses of the scalars of the model and
  estimate future sensitivities in the high-luminosity phase of the
  LHC. The searches we discuss can serve to constrain also many other
  1-loop neutrino mass models.

}

\begin{document}

\maketitle

\tableofcontents

\section{Introduction}\label{sec:introduction}

\begin{figure}[t!]
    \centering
    \includegraphics[scale=0.55]{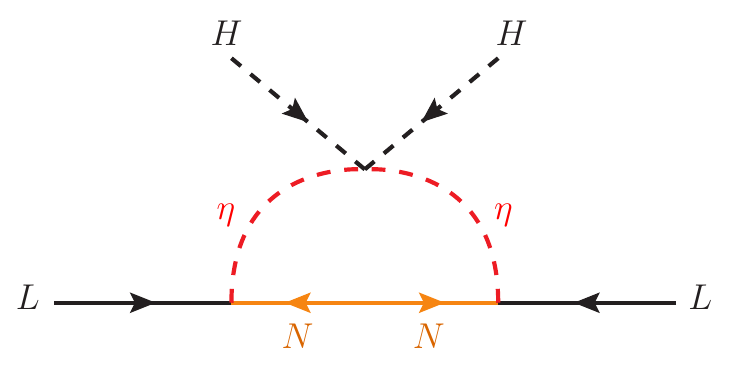}\hskip4mm
    \includegraphics[scale=0.55]{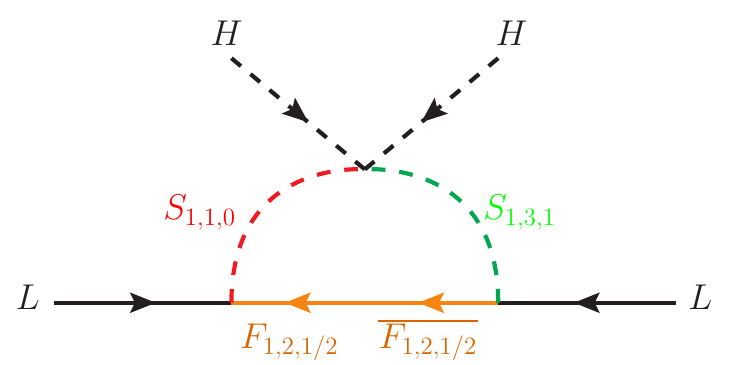}
    \caption{The Weinberg operator generated at 1-loop level in two
      example dark matter models. To the left: The original scotogenic
      model ($\eta=S_{1,2,1/2}$, $N=F_{1,1,0}$), to the right the
      singlet-triplet model that we will study in detail in this
      work.}
    \label{fig:Diags}
\end{figure}

The main idea of the famous scotogenic model \cite{Tao:1996vb,Ma:2006km}
is that neutrino masses and dark matter (DM) might share a common
origin. The model is rather simple: It adds a second Higgs doublet,
$\eta$, and three right-handed neutrinos, $N_{i}$, to the SM
particle content.  In addition, these new fields are assumed to be odd
under a discrete $Z_2$ symmetry.\footnote{The new scalar has also to
  be inert, i.e.  $\langle \eta^0\rangle=0$, which is easy to
  arrange.} The lightest $Z_2$ odd particle, either $N_{1}$ or the
lightest neutral component in $\eta$, can be a good DM candidate,
while neutrino masses are generated via a 1-loop diagram, see
fig. \ref{fig:Diags}, left.

The phenomenology of the scotogenic model has been studied in many
papers. For the case of $\eta^0$ being the lightest $Z_2$-odd
particle, the DM phenomenology of the scotogenic model resembles
closely the inert doublet model \cite{LopezHonorez:2006gr}.  Fermionic
DM and lepton flavour violation bounds on model parameters have been
calculated in \cite{Vicente:2014wga}.  A very light $N_{1}$ in the
scotogenic model could be a FIMP (``feebly interacting massive
particle''), see for example
\cite{Molinaro:2014lfa,Hessler:2016kwm,Baumholzer:2019twf}.  Indirect
detection of the DM in the scotogenic model was discussed in
\cite{deBoer:2021pon,Eisenberger:2023tgn}.  The collider phenomenology
for the case of a real scalar dark matter candidate was discussed in
\cite{Avila:2019hhv,Avila:2021mwg}. For the expectations for the
scotogenic model at a future muon collider see \cite{Liu:2022byu}.

A number of variations of the scotogenic model have also been discussed.
Examples are the ``scoto-singlet'' model (adding a $S_{1,1,0}$)
\footnote{We will use $S$ ($F$) for a scalar (fermion) BSM field, with
  the subscripts denote the properties of the field under the SM gauge
  group $SU(3)_C\times SU(2)_L \times U(1)_Y$.}
\cite{Beniwal:2020hjc} or the type-I/III variant (adding a fermionic
triplet, $F_{1,3,0}$) \cite{Hirsch:2013ola}. A version with two inert doublet
scalars (and only one $N$) was presented in \cite{Hagedorn:2018spx}. A
generalization of the scotogenic model to an arbitrary number of
generations (both scalars and fermions) was discussed in
\cite{Escribano:2020iqq}. Adding a $Z'$ to the model allows to
motivate the $Z_2$ (which is put by hand in the orignal model) and
affects drastically the relic density of the $N$ in the scotogenic
model, as discussed in \cite{Kanemura:2011vm}. A general list of UV
extensions that could explain the $Z_2$ of the scotogenic model was
given in \cite{Portillo-Sanchez:2023kbz}.  Spontaneous breaking of
lepton number in a scotogenic setup was considered in
\cite{DeRomeri:2022cem}.  All these variations keep $\eta$
and $N_{i}$ as the basic ingredient of the models. However,
there is also \cite{Lineros:2020eit}, which uses two scalar
triplets instead of $\eta$ and \cite{Farzan:2010mr} with a 
scalar triplet and a scalar singlet. 
 
On the other hand, there are many possible 1-loop neutrino mass
models. A general classification of genuine 1-loop neutrino mass
diagrams can be found in \cite{Bonnet:2012kz}. A list of possible,
phenomenologically consistent realizations has been given in
\cite{Arbelaez:2022ejo}: There are in principle two classes of
models. First, there are models in which the lightest loop particle is
either a SM field or can decay to some SM fields; these are called
``exit'' models in \cite{Arbelaez:2022ejo}. The second class contains
some BSM multiplet whose lightest member is electrically neutral and
thus can be a good cold dark matter candidate. The list of such DM
models in \cite{Arbelaez:2022ejo} is restricted to multiplets with
hypercharge at most $Y=1$, such that stringent constraints from direct
detection experiments can be avoided
\cite{Bottaro:2021snn,Bottaro:2022one}. With this condition
\cite{Arbelaez:2022ejo} counts a total 318 of DM models.

One can consider the scotogenic model as the {\em proto-type} model
for the DM model class. However, all of the DM models can explain the
observed \cite{Planck:2018vyg} relic density of dark matter, obey
bounds from direct detection (DD) experiments
\cite{LZ:2022lsv,XENON:2023cxc}, fit neutrino oscillation data
\cite{deSalas:2020pgw} and fulfill limits from charged lepton flavour
violation searches \cite{Toma:2013zsa,Vicente:2014wga}. Thus, it
seems, collider observables are the only possible way to distinguish
between (at least some of the) DM model variants.
\footnote{A disclaimer is needed here. If a signal in direct detection
  experiment were found, the mass of the WIMP, $m_{DM}$ could be
  determined, if $m_{DM}$ is small. However, for WIMP masses larger
  than the target nucleus mass, the recoil spectrum becomes nearly
  independent of $m_{DM}$ and only a lower limit on $m_{DM}$ can be
  given in this case \cite{Shan:2009ym,McDermott:2011hx}.}

One particular example model from \cite{Arbelaez:2022ejo} is shown in
fig. \ref{fig:Diags}, right. We will call this the scalar singlet-triplet (SST)
model below. Clearly, the presence of the scalar triplet, $S_{1,3,1}$,
will give rise to signals potentally very different from the
scotogenic model. The aim of our present work is to study the LHC
phenomenology of the SST model in some detail. As discussed 
in section \ref{sec:conclusions}, the signals we consider in this 
paper will appear in many other models in the list of models 
from \cite{Arbelaez:2022ejo}, too.  

The rest of this paper is organized as follows. In section
\ref{sec:Model} we present the model variant, that we will study in
detail. Then, in section \ref{sec:DarkMatter} we discuss briefly dark
matter phenomenology of the model. This study allows us to identify
parameter regions, where the model gives the correct relic density and
obeys the current bound from DD experiments. With hindsight, we
calculate the decay lengths ($c\tau$) of the doubly charged member of
the scalar triplet, $t^{++}$, and find that (i) the relic density
calculation favours points with moderate to small mass splitting
between the scalar singlet and triplets and (ii) in this mass region
parameter points that pass {\em future} direct detection bounds have
usually tiny decay widths for the $t^{++}$.  Section
\ref{sec:ColliderPhenomenology} then forms the main body of this
work. Different possible searches for $t^{++}$ at the LHC are discussed, 
current and future sensitivities are estimated. 

Finally, in section \ref{sec:conclusions} we summarize the current
paper and give also a qualitative discussion, how our results can be
applied to other 1-loop DM model variants listed in
\cite{Arbelaez:2022ejo}. Given the large number of possible models,
this discussion necessarily has to be rather superficial, but it
serves to list possible signals that would allow to distinguish many
models from the proto-type scotogenic model.

\section{Setup: Model and implementation} \label{sec:Model}

\subsection{Model lagrangian and mass matrices}\label{subsec:mdl}

The model we study in detail in this paper introduces two new scalars 
and three vector-like fermions. The quantum numbers of all new 
fields are given in table \ref{tab:qn}.



\begin{table}[h]
\centering
     \begin{tabular}{l|cccc}
              \hline \hline
      Fields &  $SU(3)_C$ & $SU(2)_L$ & $U(1)_Y$ & $ \mathbb{Z}_2 $ \\ 
\hline
      (Real) $S_{1}$ & $1$ & $1$ & $0$ & $-$\\
      $S_{3}$ & $1$ & $3$ & $1$ & $-$ \\ 
      $F$ $(\overline{F})$& $1$ & $2$ & $1/2$ $(-1/2)$ & $-$ \\ \hline \hline
\end{tabular}
\caption{Quantum numbers for the particle content of the model. 
While neutrino masses could be explained already with one copy 
of $F/\overline{F}$, in our numerical study we use three families of 
$F/{\bar F}$ to reflect the number of generations in the SM.}
\label{tab:qn}
\end{table}

The triplet scalar can be written in the usual way:
\begin{equation}
S_{3}=\begin{pmatrix}
 t^{+}/\sqrt{2} & t^{++}\\
t^{0} & -t^{+}/\sqrt{2} 
\label{Eq:}
\end{pmatrix}.
\end{equation}
The Lagrangian of the model adds the following terms to 
the SM Lagrangian:
\begin{eqnarray}\label{eq:lag}
  {\cal L} = 
  &-& (M_F F{\bar F} +  M_{S_1}^2 |S_{1}|^2 + M_{S_3}^2 |S_{3}|^2 )
   -  ( h_F L FS_{1}+ h_{\bar F} L {\bar F}S_{3} + {\rm h.c.})
  \\ \nonumber
  & + & \lambda_2 |H|^2 |S_{1}|^2 
  +   \Big(\lambda_{3a}|H|^2 |S_{3}|^2
  + \lambda_{3b} \epsilon_{l_1l_3}\epsilon_{l_2l_4}\epsilon_{l_{3b}l_{4_b}}
  H_{l_1}H^*_{l_2}(S_{3})_{l_3l_{3b}}(S_{3}^*)_{l_4l_{4b}}\Big)
  \\ \nonumber
  & + &   \lambda_4 (|S_1|)^4 - ( \lambda_5 HHS_1 S_3^{\dagger} + {\rm h.c})
  \\ \nonumber
  & +& \Big( \lambda_{6a} (|S_3|)^4  +\lambda_{6b}
  \epsilon_{l_1l_3}\epsilon_{l_{1b}l_{3b}}\epsilon_{l_{2}l_{4}}\epsilon_{l_{2b}l_{4_b}}
  (S_3)_{l_1l_{1b}}(S_3^*)_{l_2l_{2b}} (S_3)_{l_3l_{3b}}(S_3^*)_{l_4l_{4b}}\Big)
  +\lambda_7 |S_1|^2|S_3|^2.
\end{eqnarray}
The terms proportional to $\lambda_3$ and $\lambda_6$ have two
independent $SU(2)_L$ contractions each. We calculated these and
cross-checked for consistency with \texttt{Sym2Int}
\cite{Fonseca:2017lem}. The couplings $\lambda_{6a}$, $\lambda_{6b}$
and $\lambda_{7}$ describe only scatterings among the BSM scalars
$S_1$ and $S_3$, for collider and dark matter phenomenology they are
mostly irrelevant and will therefore not be discussed in detail in the
rest of this paper.  $M_{F}$, in general, is a $3 \times 3$ matrix. We
can always choose a basis, such that $M_F$ is diagonal and real, with
eigenvalues $m_{F_i}$. The Yukawa matrices $h_{F}$, $h_{\bar F}$ are
$3 \times 3$ non-diagonal and can be complex in general. From the
quartic couplings, the most important one is $\lambda_5$. In the limit
$\lambda_5\to 0$, the model conserves lepton number and neutrino
masses are zero.

After electroweak symmetry breaking the neutral part of the scalars in
$S_{1}$ and $S_{3}$ mix.  Their (tree-level) mass matrix is given by:
\begin{equation}
\large
M_{S}^{0}=\begin{pmatrix}
 M_{S_{3}}^2-\frac{\lambda_{3a} v^2}{2} & \frac{\lambda_{5} v^2}{\sqrt{2}} \\
 \frac{\lambda_{5} v^2}{\sqrt{2}}  &  M_{S_{1}}^2-\frac{\lambda_{2} v^2}{2} 
 \label{Eq:2}
\end{pmatrix},
\end{equation}
The eigenvalues of this matrix can be easily calculated:
\begin{eqnarray}
(m_{s^{0}_1,s^{0}_2})^{2} & = &\frac{1}{4} \Big( 2 (M_{S_1}^{2}+ M_{S_3}^{2})
            -(\lambda_{2}+\lambda_{3_{a}})v^2 
\\ \nonumber
&\mp & \sqrt{\big(2 (M_{S_1}^{2} -M_{S_3}^2)
  +v^2 (\lambda_{3_{a}}-\lambda_{2})\big)^2 + 8 \lambda_{5}^2  v^4}\Big),
\end{eqnarray}
The mixing angle between the neutral states is given by:
\begin{equation}
  \tan 2\theta =  \frac{\sqrt{2}\lambda_5 v^2}
      {( M_{S_3}^2- M_{S_1}^2) + \frac{1}{2}( \lambda_2 - \lambda_{3a} ) v^2}.
  \label{eq:t2t}
\end{equation}
This mixing angle between $t_0$ and $S_0$ is usually small (unless
$M_{S_1}^2 \equiv M_{S_3}^2 - v^2 (\lambda_{3a}-\lambda_{2})/2$, where
the denominator of eq. (\ref{eq:t2t}) becomes zero), since typically
$\lambda_5 v^2 \ll M_{S_1}^{2},M_{S_3}^2$. For $M_{S_1} \le M_{S_3}$
the lighter of the neutral scalars, $s^{0}_1$, is mostly a singlet
while $s^{0}_2 \simeq t^0$. In a slight abuse of notation, we will
call the lightest neutral scalar, $s^{0}_1$, simply $S_0$ from now on.

On the other hand, the masses of the charged components of the scalar
triplet $t^{++}$ and $t^{+}$ are:
\begin{eqnarray}\label{eq:msmp}
  m_{t^{++}}^2 & = \ M_{S_3}^2 -\frac{1}{2} \left ( \lambda_{3a}+\lambda_{3b} \right )v^2,
  \\ \nonumber
   m_{t^{+}}^2 & = \ M_{S_3}^2 - \frac{1}{4} \left ( 2\lambda_{3a}+\lambda_{3b} \right )v^2.
   \\ \nonumber
\end{eqnarray}
We will be interested in studying DM in this model, thus typically 
the lightest neutral state will be mostly singlet with some (small) 
admixture from $t_0$.\footnote{A pure $t_0$ is not a good DM candidate,
  due to constraints from direct detection experiments~\cite{LZ:2022lsv}.}

For completeness, we mention that due to the $Z_2$ symmetry the 
vector-like fermions do not mix with the SM leptons. The charged 
and the neutral member of $F$ are not exactly degenerate, since 
1-loop corrections from QED affect the mass of $F^+$ but not 
$F^0$. This mass splitting is of order $0.3$ GeV \cite{Cirelli:2005uq}, 
sufficiently large that the decay $F^+ \to \pi^+ + F^0$ is always 
kinematically possible.

The model generates 1-loop neutrino masses from diagram T-3 of
\cite{Bonnet:2012kz}. The neutrino mass matrix is given by:
\begin{equation}\label{eq:numass}
  m_{\nu} = {\cal F} \Big( h_F^T {\hat M}_R h_{\bar F}
             +  h_{\bar F}^T {\hat M}_R h_{F} \Big).
\end{equation}
Here ${\cal F} \sim \frac{1}{2}\frac{\sin 2 \theta}{16 \pi^2}$ and
${\hat M}_R$ is a $(3 \times 3)$ matrix with diagonal entries given
by:
\begin{equation}\label{eq:MR}
  ({\hat M}_R)_{ii} = \Delta B(x_i,y_i) m_{F_i}, 
\end{equation}
with $x_i=((m_{s^{0}_1}/m_{F_i})^2$, $y_i=((m_{s^{0}_2}/m_{F_i})^2$,
$m_{s_{i}}^{0}$ the eigenvalues of Eq.\ref{Eq:2} and
\begin{equation}\label{eq:MR}
  \Delta B(x,y) = \frac{x\log(x)}{x-1} - \frac{y\log(y)}{y-1} .
\end{equation}
As shown in eq. (\ref{eq:numass}) the neutrino masses depend on the
product of two $h_{F}$, $h_{\bar F}$ Yukawa couplings, which in
principle can be different. Fits to neutrino data of the parameters in
eq. (\ref{eq:numass}) can be done using the master parametrization
\cite{Cordero-Carrion:2019qtu}. Since we will not study flavour in
detail in this paper, we use a simplified version of the general
formula in our fits. This simplification does not affect the decay
length nor the DM calculation. In these fits, we express $h_{F}$ in
terms of the measured neutrino masses and angles, the fermion mass
eigenvalues $m_{F_i}$ and $h_{\bar F}$, which we choose randomly,
resulting in:
\begin{equation}\label{eq:hf}
 h_{F} = \frac{1}{2 {\cal F}}{\hat M}_{R}^{-1} (h_{\bar F}^{T})^{-1} 
          U_{\nu}^* \hat{m_{\nu}} U_{\nu}^{\dagger} ,
\end{equation}
where $[M_{R}] = [{\rm GeV}]$, $[m_{\nu}] = [{\rm eV}]$, while the Yukawa couplings $h_{F}$ and $h_{\bar
  F}$ are adimensional. Here, $\hat{m_{\nu}}$ are the light neutrino
mass eigenvalues ($m_{\nu_{i}}={m_{\nu_{1}}}$,
$\sqrt{\Delta(m_{sol}^2)+m_{\nu_{1}}^2}$,
$\sqrt{\Delta(m_{atm}^2)+m_{\nu_{1}}^2}$), with $\Delta(m_{sol}^2)$
and $\Delta(m_{atm}^2)$  the solar and atmospheric mass squared
splittings and $U_{\nu}$ is the neutrino mixing matrix, measured in
oscillation experiments.

We note that the values of $\lambda_{5}$ that we use in our numerical
analysis are compatible with neutrino oscillation data and also with
current upper limits on charged lepton flavor violating (cLFV) decays,
such as $\mu \rightarrow e \gamma$. Avoiding significant cLFV
observables can be achieved in different ways. The two simplest
possibilities are: i) Choosing a large enough value of $\lambda_{5}$,
which, see eq. (\ref{eq:numass}), results in $h_{F} \ll 1 $ and
$h_{\bar F} \ll 1 $ in the neutrino fit. In this scenario, the model
clearly satisfies cLFV constraints. ii) Choosing either $h_{F}$ or
$h_{\bar F}$ to be diagonal and order ${\cal O}(1)$. This then
automatically leads to non-diagonal yukawa couplings $\ll 1$, thus
cLFV constraints are fulfilled again.  In our model, cLFV current
bounds are always irrelevant for values of $\lambda_{5}$ larger than
$\lambda_{5} > 10^{-6}$. In our numerical analysis, we do not consider
values of $\lambda_{5}$ smaller than this value.

\subsection{Model implementation and decay widths}\label{subsect:decw}

We implemented the model into SARAH \cite{Staub:2013tta,Staub:2012pb}.
SARAH automatically generates \texttt{SPheno} routines
\cite{Porod:2003um,Porod:2011nf} for the numerical calculation of the
mass spectra, mixing matrices, scalar and fermionic two-body decays,
and other observables. SARAH also generates model files for
\texttt{MicroMegas} \cite{Belanger:2013oya}, with which we calculate
the relic density and the direct detection cross section for our dark
matter candidate. UFO input files for \texttt{MadGraph}
\cite{Alwall:2007st,Alwall:2011uj} can also be generated, which we
then use for the calculation of 3-body, 4-body and 5-body decay widths
of the exotic scalars, as well as for the production cross sections.

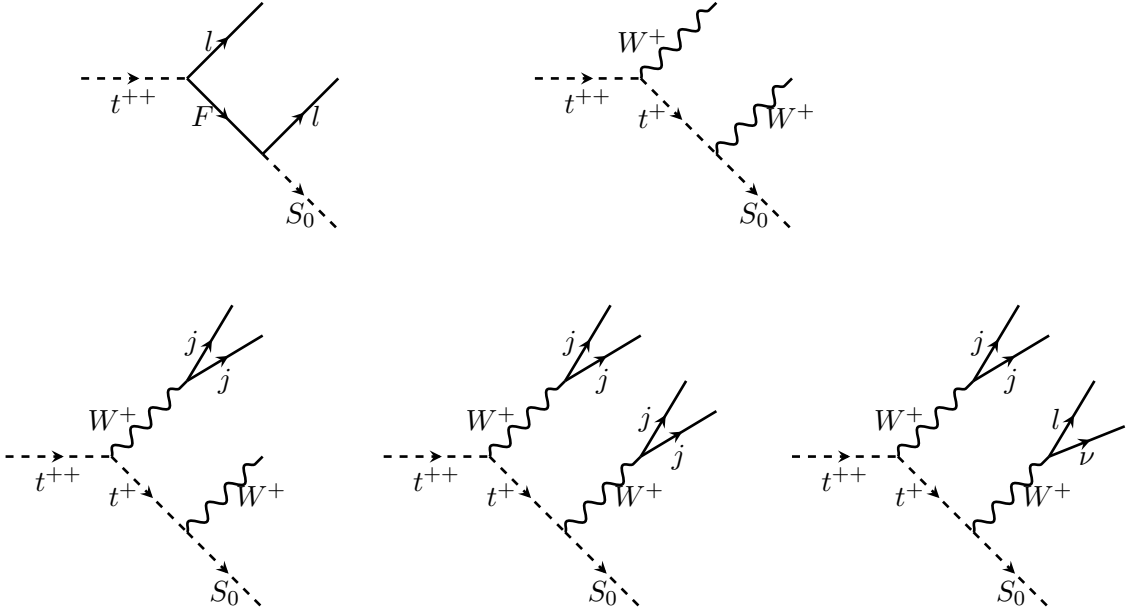
\begin{figure}[H]
    \centering
    \begin{tikzpicture}[line width=1. pt, scale=2]
        \draw[scalar] (-1.7+0.5,0) -- (-1+0.5,0)
            node[midway, below] {$t^{++}$};

             \draw[fermion] (-1+0.5,0) -- (-0.5+0.5,0.5)
            node[midway, left] {$l$};
        
        \draw[fermion] (-1+0.5,0) -- (-0.5+0.5,-0.5)
            node[midway, left] {$F$};

             \draw[fermion] (-0.5+0.5,-0.5) -- (0+0.5,0)
            node[midway, right] {$l$};

             \draw[scalar] (-0.5+0.5,-0.5) -- (0+0.5,-1)
            node[midway, below] {$S_{0}$};
\draw[scalar] (1.8,0) -- (2.5,0)
            node[midway, below] {$t^{++}$};

             \draw[vector] (2.5,0) -- (3,0.5)
            node[midway, left] {$W^{+}$};
        
        \draw[scalar] (2.5,0) -- (3,-0.5)
            node[midway, left] {$t^{+}$};

             \draw[vector] (3,-0.5) -- (3.5,0)
            node[midway, right] {$W^{+}$};

             \draw[scalar] (3,-0.5) -- (3.5,-1)
            node[midway, below] {$S_{0}$};


          \draw[scalar] (-0.7-1,-2.5) -- (0-1,-2.5)
            node[midway, below] {$t^{++}$};

             \draw[vector] (0-1,-2.5) -- (0.5-1,-2)
            node[midway, left] {$W^{+}$};

            \draw[fermion] (-0.5,-2) -- (-0.2,-1.5)
            node[midway, left] {$j$};

            \draw[fermion] (-0.5,-2) -- (0,-1.7)
            node[midway, below] {$j$};

        \draw[scalar] (0-1,-2.5) -- (0.5-1,-3)
            node[midway, left] {$t^{+}$};
        
             \draw[vector] (0.5-1,-3) -- (1-1,-2.5)
            node[midway, right] {$W^{+}$};    

             \draw[scalar] (0.5-1,-3) -- (1-1,-3.5)
            node[midway, below] {$S_{0}$};


             \draw[scalar] (0.3+0.5,-2.5) -- (1+0.5,-2.5)
            node[midway, below] {$t^{++}$};

             \draw[vector] (1+0.5,-2.5) -- (1.5+0.5,-2)
            node[midway, left] {$W^{+}$};

           \draw[fermion] (2,-2) -- (2.3,-1.5)
            node[midway, left] {$j$};

             \draw[fermion] (2,-2) -- (2.5,-1.7)
            node[midway, below] {$j$};

        \draw[scalar] (1+0.5,-2.5) -- (1.5+0.5,-3)
            node[midway, left] {$t^{+}$};

             \draw[vector] (1.5+0.5,-3) -- (2+0.5,-2.5)
            node[midway, right] {$W^{+}$};

 \draw[fermion] (2.5,-2.5) -- (2.8,-2)
            node[midway, left] {$j$};

            \draw[fermion] (2.5,-2.5) -- (3,-2.2)
            node[midway, below] {$j$};

             \draw[scalar] (1.5+0.5,-3) -- (2+0.5,-3.5)
            node[midway, below] {$S_{0}$};   


         \draw[scalar] (3.5,-2.5) -- (4.2,-2.5)
            node[midway, below] {$t^{++}$};

             \draw[vector] (4.2,-2.5) -- (4.7,-2)
           node[midway, left] {$W^{+}$};

 \draw[fermion] (4.7,-2) -- (5,-1.5)
           node[midway, left] {$j$};

            \draw[fermion] (4.7,-2) -- (5.2,-1.7)
           node[midway, below] {$j$};

        \draw[scalar] (4.2,-2.5) -- (4.7,-3)
            node[midway, left] {$t^{+}$};

             \draw[vector] (4.7,-3) -- (5.2,-2.5)
            node[midway, right] {$W^{+}$};

              \draw[fermion] (5.2,-2.5) -- (5.5,-2.0)
            node[midway, left] {$l$};

\draw[fermion] (5.2,-2.5) -- (5.7,-2.3)
            node[midway, below] {$\nu$};

             \draw[scalar] (4.7,-3) -- (5.2,-3.5)
            node[midway, below] {$S_{0}$};   
       
    \end{tikzpicture}
    \caption{Example Feynman diagrams for $t^{++}$ decays. The
      diagrams in the first line show 3-body decays, while the
      diagrams in the bottom line show the corresponding 4- and 5-body decays,
      if the mass splitting between the $t^{++}$ and the dark matter cadidate
    is smaller than $2m_W$ and $m_W$, respectively.}
    \label{fig:tripletdecays}
\end{figure}

With hindsight of the results in sections \ref{sec:DarkMatter} and
\ref{sec:ColliderPhenomenology}, we will next briefly describe scalar
decay widths. Here, we will assume that the fermions have masses
larger than $t^{++}$ and $m_{t^{++}}-m_{t^{+}} \le m_W$, such that
there are no two-body decay channels allowed kinematically for
$t^{++}$. The doubly charged component of the scalar triplet will
decay to the lightest neutral scalar, usually mostly a scalar singlet
(and the dark matter candidate of the model), and either two charged
leptons or two $W^+$-bosons, via the diagrams shown in the top line
of fig.  \ref{fig:tripletdecays}.
If the mass splitting, $\Delta m \equiv m_{t^{++}}-m_{S_0}$, is smaller than
$2m_{W}$ ($m_W$), at least one of the $W^+$s will be off-shell and the
decay $t^{++}\to W^+ + (W^+)^* + S_0$ becomes effectively a 4-body
(5-body) decay, as shown in the diagrams in the second line of
fig. \ref{fig:tripletdecays}.

The $W^+$ themselves can either decay leptonically or hadronically,
thus the 5-body final states can be either $4j+S^0$, $2j l\nu+S^0$ or
$2l2\nu+S^0$. We define $\Gamma_{ll} = \Gamma(t^{++} \rightarrow l^{+}
+ l^{+} + S_{0})$ and $\Gamma_{had}=\Gamma (t^{++} \rightarrow 4j +
S_{0} ) + \Gamma (t^{++} \rightarrow 2j + l \nu + S_{0} )$. Note that
the final state $2l2\nu+S^0$ (mediated by $W$s) can not be easily
distinguished from $2l+S^0$, due to the missing energy (from the
$S^0$) in the events.

\begin{figure}[ht]
	\centering
	\includegraphics[width=0.49\textwidth]{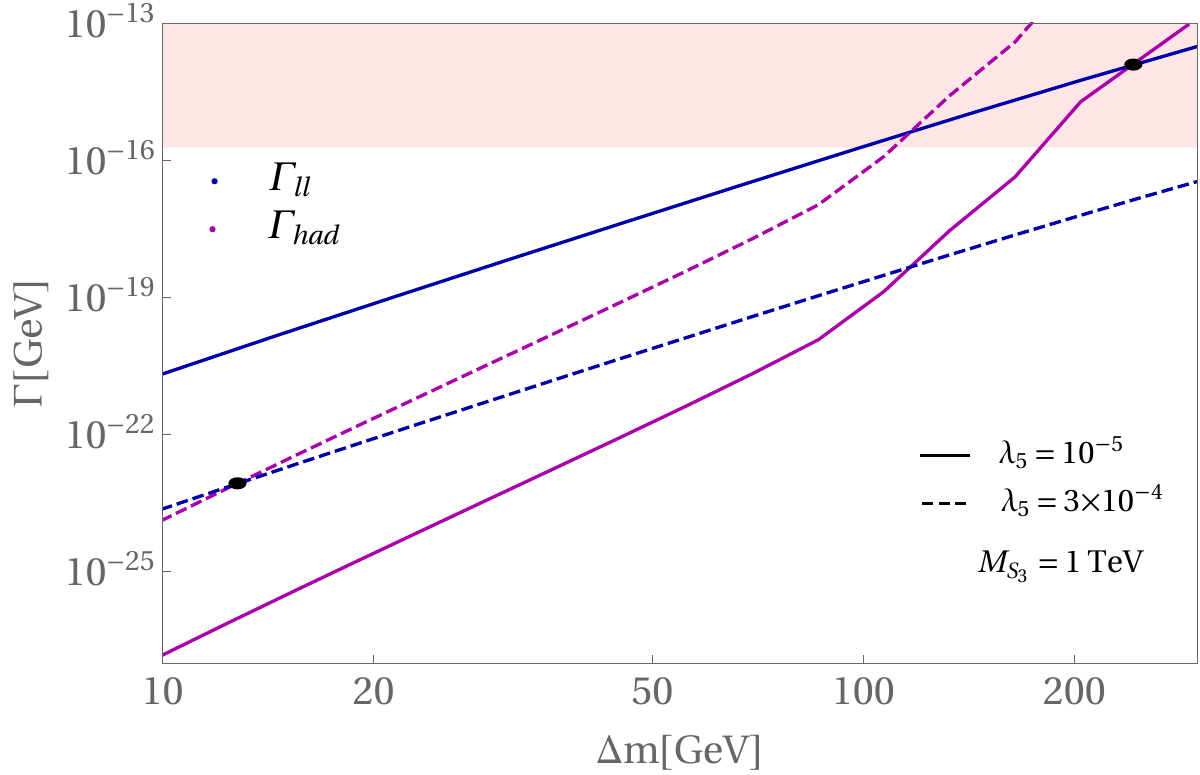}
	\includegraphics[width=0.49\textwidth]{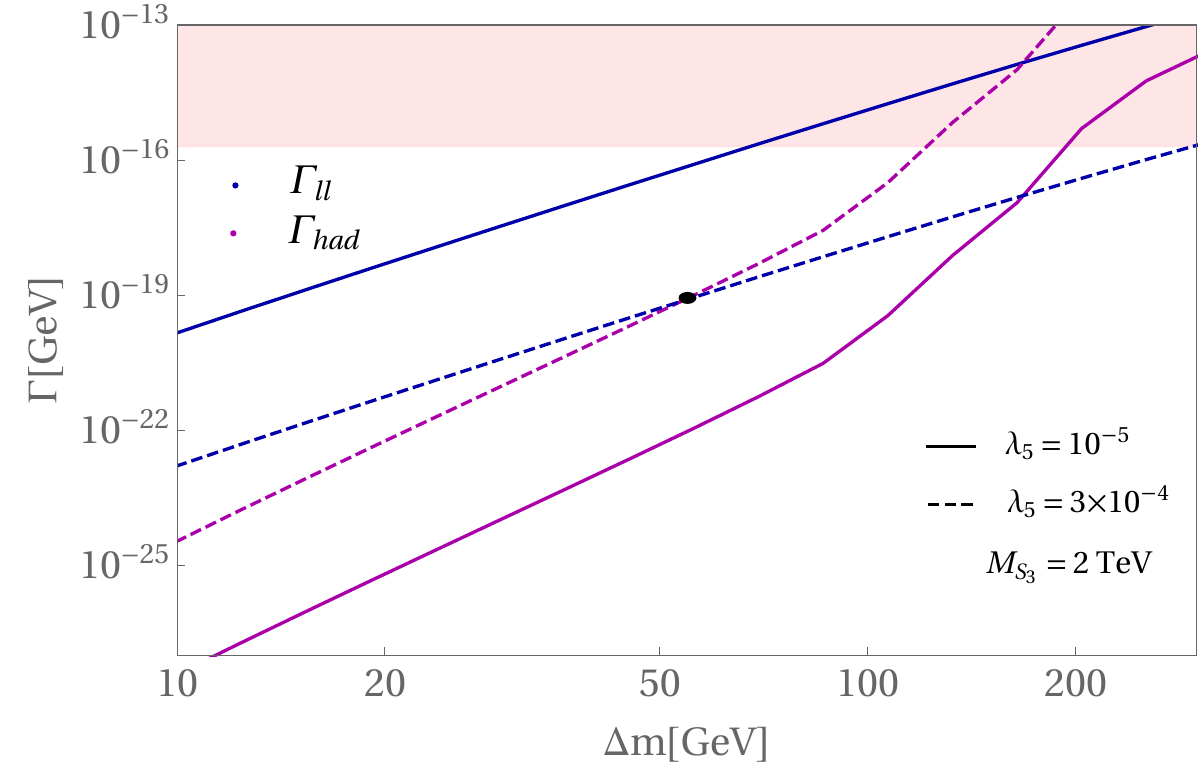}
	\caption{Decay width as a function of $\Delta m$ for
          different values of $\lambda_{5}$. The left (right) panels
          correspond to $M_{S_{3}} = 1$ TeV ($M_{S_{3}} = 2$ TeV). The
          amplitudes $\Gamma_{ll}$ and $\Gamma_{had}$ refer to $\Gamma
          (t^{++} \rightarrow l^{+} + l^{+} + S_{0} )$ and $\Gamma
          (t^{++} \rightarrow 4j + S_{0} ) + \Gamma (t^{++}
          \rightarrow 2j + l\nu + S_{0} )$ respectively. The pink band
          corresponds to the region where $c\tau$ of $t^{++}$
          corresponds to the ``displaced'' regime, i.e $c\tau \sim$ (mm
          - m).  For this figure we have fixed $m_{F} = 2.5$ TeV.}
   \label{fig:crossingpoints}
\end{figure}

For the collider phenomenology the total decay length of $t^{++}$ is
important. We can essentially distinguish three regimes: prompt decays
($c\tau \lesssim {\cal O}(1)$ mm), finite lengths, $c\tau \sim$ (mm - m), 
or quasi-stable events ($c\tau \gtrsim {\cal O}(10)$  m). In fig.
\ref{fig:crossingpoints} we show some example decay widths of
$t^{++}$ as a function of $\Delta m$, for different
choices of $\lambda_5$ and $M_{S_3}$. Note that in this figure
we always fit neutrino masses such that they reproduce oscillation
data (for normal hierarchy). This leads to a fixed choice of
$h_{F} h_{\bar{F}}$, once $\lambda_5$ and $m_F$ are fixed.

As fig. \ref{fig:crossingpoints} demonstrates, which of the final
states, ``hadronic'' (whose width we define as $\Gamma_{had}$) or ``leptonic'' (whose width we define as $\Gamma_{ll}$), dominates the decay width
depends mostly on the choice of $\lambda_5$ -- and to a lesser extend
-- on $\Delta m $ and $M_{S_3}$.\footnote{There is not much dependence
on $m_{F}$, since a choice of a larger $m_{F}$ is partially
compensated by a larger $h_{F} h_{\bar{F}}$ from the neutrino mass
fit.} Generically, for $\lambda_5$ larger than (few) $10^{-4}$ the
hadronic channel will dominate the decay of $t^{++}$, regardless of
other model parameter choices, unless $\Delta m $ is very small (say,
$\Delta m  \simeq$ (few) GeV), where a strong phase space suppression
for the 5-body decays occurs. The exact value of $\Delta m$, where
$\Gamma_{had}$ becomes larger than $\Gamma_{ll}$ depends on $M_{S_3}$
of course, but larger $\Delta m $ prefers the hadronic
channels (since $\Gamma_{had}$ has a stronger dependence on
$\Delta m$ than $\Gamma_{ll}$). 

Very generally, the decay width of $t^{++}$ is expected to be a very
small number, essentially because the smallness of neutrino masses
require that the product $\lambda_5h_{F} h_{\bar{F}}$ is small.
Note that large part of the parameter space shown in fig.
\ref{fig:crossingpoints} actually leads to widths that are so small
that the $t^{++}$ becomes quasi-stable. In this case the $t^{++}$ will
leave a track in the LHC detectors, but the final states are not
accesible, of course.

\section{Dark matter phenomenology}\label{sec:DarkMatter}

The presence of a preserved $Z _2$ symmetry in the SST model implies
that the lighest $Z _2$-odd particle is completely stable, hence a
potential DM candidate.  Among the possible fermionic and scalar
particles, we focus on the scenario where $S _0$ is the lightest
$Z_2$-odd state. As aforementioned, $S _0$ is a CP-even mass
eigenstate resulting from the mixing between the neutral components of
the singlet and triplet scalars. Under the assumption $M _{S _3} ^2 >
M _{S _1} ^2$, $S _0$ will typically be the lightest scalar for most
values of the quartic couplings $\lambda _i$, provided they remain
perturbative. Moreover, if $\lambda _5$ is small, the $S _1$-$S _3$
mixing is supressed and $S _0$ will be predominantly the real scalar
singlet $S _1$. Therefore, it inherits some properties of singlet
scalar DM candidates \cite{Arcadi:2019lka, Cline:2013gha,
  Bhattacharya:2019fgs, DuttaBanik:2020jrj}. Additionally, we assume
that $m _F ^2 > M _{S _1} ^2$, ensuring that the vectorlike leptons
remain havier than $S _0$. Throughout our analysis, we enforced these
conditions and considered points in the parameter space where $S _0$
is the lightest $Z _2$-odd particle.

The $S _0$ relic abundance $\Omega h ^2$ can be generated within the
standard thermal WIMP paradigm, leading to \cite{Griest:1990kh}
\begin{equation}
\Omega h ^2 = \frac{1.09 \times 10 ^9 \text{GeV} ^{-1}}{g _* ^{1/2} M _{PL}} \frac{1}{J(x _f)} ,
\end{equation}
where $g _*$ denotes the number of relativistic degrees of freedom at
freeze-out, $x _f = m _{S _0} / T _f$ and
\begin{equation}
J(x) = \int _{x} ^\infty \frac{\langle \sigma v \rangle}{x ^2} dx ,
\end{equation}
with $\langle \sigma v \rangle$ being the thermal average of the DM
annihilation cross section. For a hierarchical spectrum of $Z _2$-odd
scalar particles, the primary annihilation channels contributing to
$\langle \sigma v \rangle$ stem typically from Higgs mediated
processes controlled by the quartic couplings, most prominently
$\lambda _2$, which couples $S _1$ to the Higgs doublet
$H$. Contributions from terms proportional to $\lambda _{3a}$ and
$\lambda _{3b}$, which link $S _3$ to $H$, are also significant,
though less relevant in the limit of small $S _1$-$S _3$ mixing. As
these quartic couplings also set the strength of DM-nucleus scattering
cross section, dominated by $t$-channel Higgs exchange, this scenario
of DM production driven by Higgs portal couplings faces strong
constraints from current direct detection searches
\cite{LZ:2022lsv,XENON:2023cxc}.

\begin{figure}[t]
\centering
\includegraphics[width=0.6\textwidth]{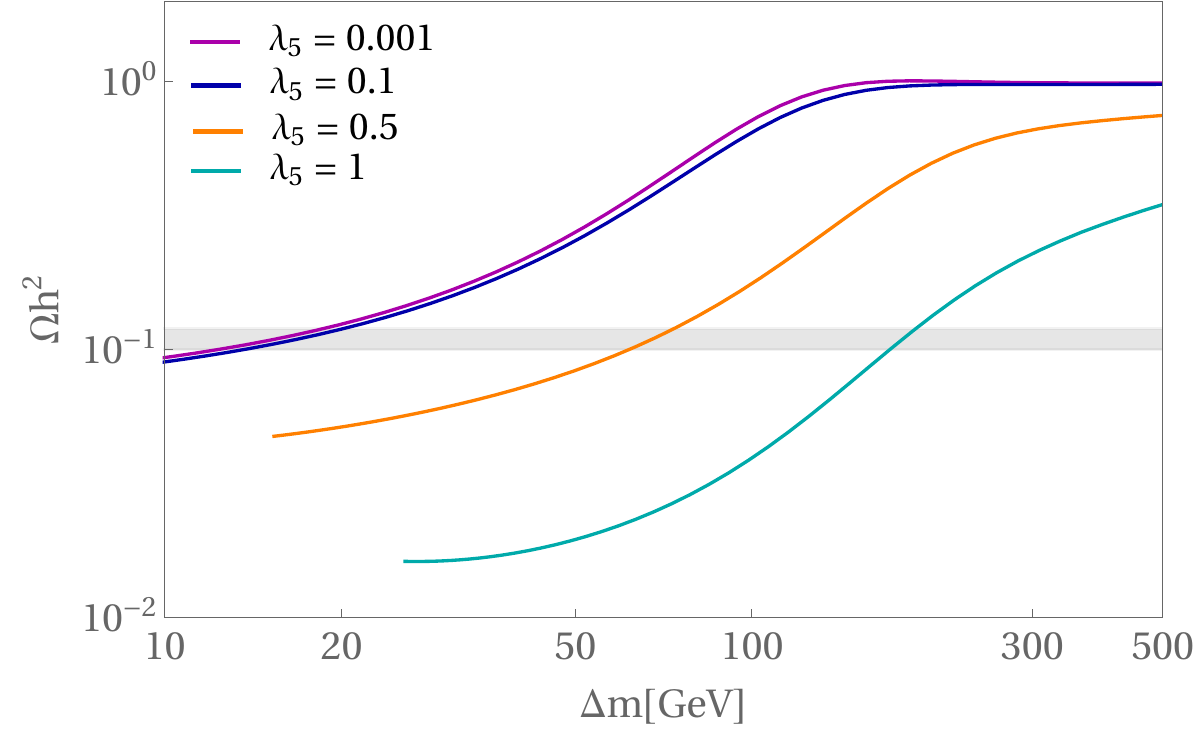}
\caption{The plot shows $\Omega h^2$ as function of $\Delta m$ for
  four different choices of $\lambda_5$ for one fixed, but arbitrary
  choice of $m_{S_{0}}= 1$ TeV. Smaller values of $\lambda_5$ lead to
  smaller mixing between the neutral singlet and triplet states and,
  thus to a larger $\Omega h^2$. However, even for very small
  $\lambda_5$, the correct $\Omega h^2$ can be achieved if $\Delta m$
  is small. The plot therefore demonstrates the importance of
  coannihilation.}
    \label{fig:DM_coann}
\end{figure}

When the mass splitting between the singlet and triplet scalars is not
significantly large, however, co-annihilation channels involving $t
^{++}$, $t ^{+}$, $t ^0$ come into play and dominate $\langle \sigma v
\rangle$. The importance of co-annihilation processes becomes
substantial when the relative mass splittings of the triplet scalars
with $S _0$ are much less than $1$ \cite{Griest:1990kh}. In this case,
$\lambda _5$ and $\Delta m$ become the key parameters in determining
the DM relic abundance. This is demonstrated in
fig. (\ref{fig:DM_coann}) for some specific example. Co-annihilation
allows the correct $\Omega h ^2$ to be obtained in a broader range of
$S _0$ masses.

\begin{figure}[t]
\centering
\includegraphics[width=0.49\textwidth]{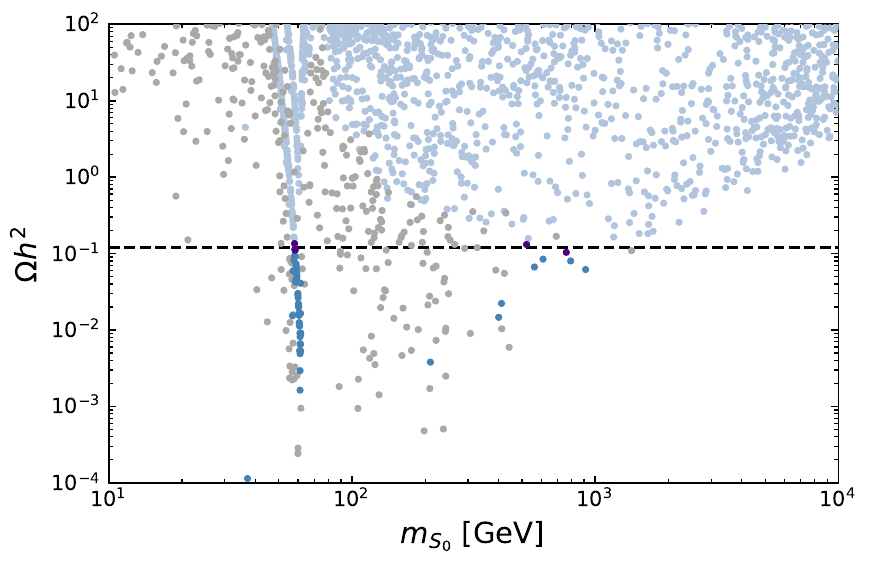}
\includegraphics[width=0.49\textwidth]{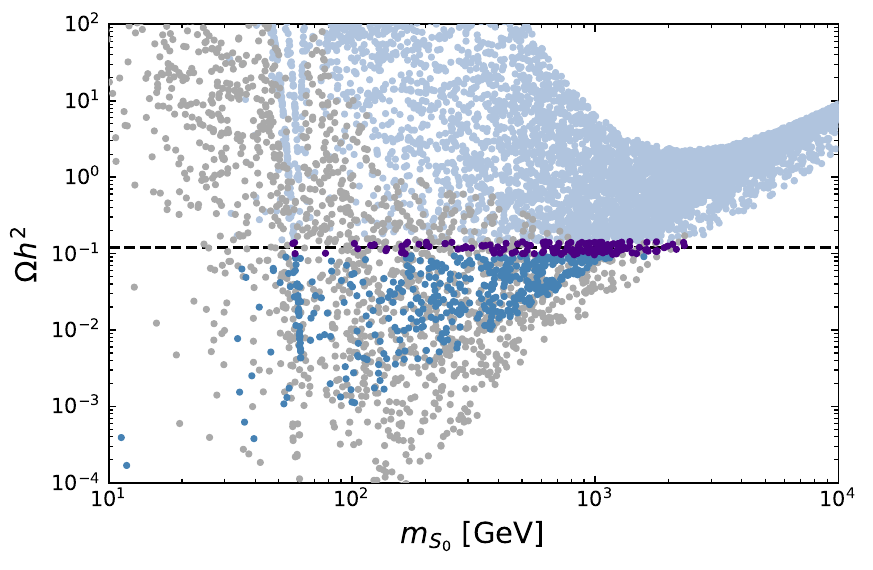}
\caption{Dark matter relic density in the non-coannihilation (left
  panel) and coannihilation (right panel) regimes (see the
  text). Violet points comply with the relic density value as measured
  by Planck. The grey points are excluded by direct detection
  constraints, while blue (light blue) points pass the direct
  detection bound, but yield underabundant (overabundant) DM. Note
  that the presence of coannihilation allows many viable points for $m
  _{S _0}$ in the range of hundreds GeV to few TeV.}
    \label{fig:DM_relic_den}
\end{figure}

In fig. \ref{fig:DM_relic_den} the DM relic abundance is calculated
for the two different scenarios discribed above, i.e., the
hierarchical scalar spectrum (left panel), and compressed scalar
spectrum (right panel). In these plots, the quartic couplings are
fixed as $\lambda _2 = \lambda _{3a} = \lambda _{3b} = 10 ^{-3}$,
while the other parameters are varied randomly in the following
ranges: $m _{S _0} \subset [ 10~\text{GeV}, 10~\text{TeV} ]$, $\lambda
_5 \subset [ 10 ^{-4}, 1 ]$, $\Delta m \subset [ 0, 5~\text{TeV} ]$
(in the left panel) and $\Delta m \subset [ 0, 200~\text{GeV} ]$ (in
the right panel). $m _F$ varies in the same range as $m _{S _0}$,
however always respecting the condition $m _F > m _{S _0}$. In the
numerical analysis, we utilized the {\texttt{MicrOmegas v5.2.4}}
package \cite{Belanger:2013oya} to compute the DM abundance and also
to compute the scattering cross section for direct detection.
Additionally, consistency with neutrino oscillation data was enforced
by the use of Eq.~\eqref{eq:hf}, for which a fixed value of $h
_{\bar{F}} = 10 ^{-2}$ has been chosen for the Yukawa couplings.

As can be seen in the left panel of fig. \ref{fig:DM_relic_den}, in
the hierarchical spectrum scenario the majority of points that
reproduce the correct relic density are excluded by LUX-ZEPLIN (LZ)
data \cite{LZ:2022lsv}, except for small regions where $m _{S _0}$ is
around $m _h / 2$ (the Higgs resonance) or within the restricted
interval between $500$-$800$ GeV. Larger $m _{S _0}$ values fail to
generate the correct $\Omega h ^2$ for the chosen fixed values of
$\lambda_{2, 3a, 3b}$. Even though we could allow heavier $S _0$ by
making $\lambda_{2, 3a, 3b}$ larger, these points would be ruled out
by direct detection, as the increase in the couplings lead to larger
DM-nucleus scattering cross-section, mediated by the Higgs. Besides,
this setting is less attractive from the phenomenological point of
view, as the extra particles tend to be heavier than TeV, making their
search at colliders a difficult task.

As the hierarchical scenario is severely constrained by direct
detection, we turn our attention to the alternative case, shown in the
right panel of fig. \ref{fig:DM_relic_den}. In this plot, the mass
splitting is restricted to be less than $200$ GeV, so most of the
points fall within the co-annihilation regime. In this situation, the
additional channels enhance the annihilation efficiency without the
need of increasing the Higgs portal couplings. Consequently, the relic
density is correctly generated over a broader range of $m _{S _0}$
values, including small masses of the order of a few hundred GeV,
while evading direct detection constraints.  Note that if $S _0$ is
light, the small mass splitting implies that the other scalars are
also light, and hence can potentially be produced at the LHC.

\begin{figure}[t]
\centering
\includegraphics[width=0.65\textwidth]{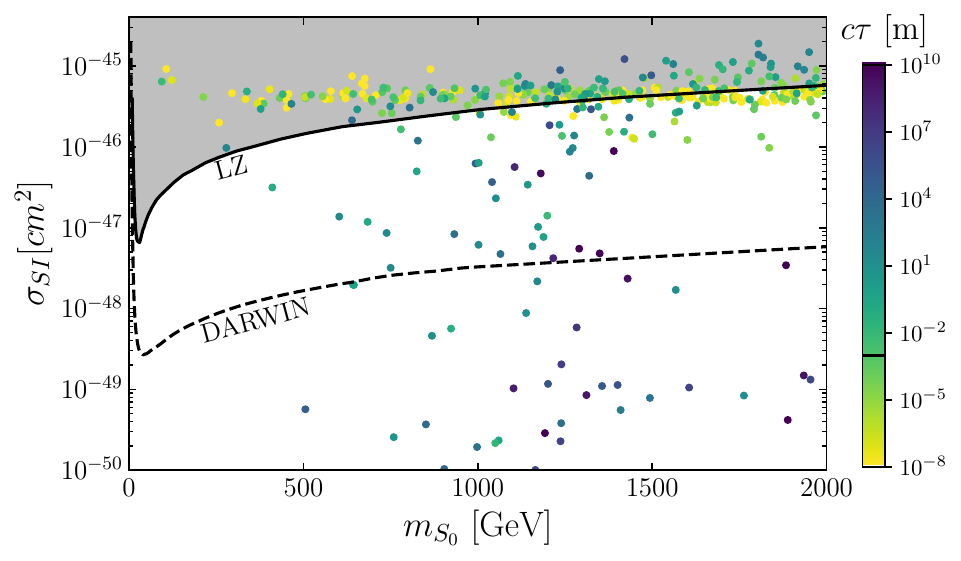}
\caption{Spin-independent DM direct detection cross section versus DM
  mass. The sidebar indicates the lifetime of the $t^{++}$ scalar,
  with bluish (yellowish) points corresponding to long-lived
  (short-lived) $t ^{++}$. The strip in the sidebar corresponds to $c
  \tau = 1$ mm, which is the threshold above which we consider a
  particle to be long-lived. The $t ^{++}$ particle is long-lived for
  the majority of points that satisfy current direct detection
  constraints, including \textit{all} the points that evade future
  detection by DARWIN.}
    \label{fig:DD_xsec_ms0}
\end{figure}

An additional feature that emerges in the co-annihilation regime,
favored by DM, is the fact that the doubly charged scalar $t ^{++}$ is
typically long-lived.  This is illustrated in
fig.~\ref{fig:DD_xsec_ms0}, where the DM direct detection cross
section is plotted against the DM mass, with the sidebar showing the
decay length of $t ^{++}$. In this scan, all the points comply with
the Planck constraint $\Omega h ^2 = 0.1200 \pm 0.0036 $ (at $3
\sigma)$ \cite{Planck:2018vyg}, and the neutrino mass fit as discribed
previously. The quartic couplings are allowed to vary in the range
$\lambda _i \subset [10 ^{-6}, 1] $, and $\Delta m$ varies from $0$ to
$500$ GeV. As $m _{S _0}$ in the horizontal axis goes up to $2$ TeV,
we fixed $m _F = 2.5$ TeV. It is apparent that the points currently
allowed by LZ bounds are mostly bluish, which means that $t ^{++}$ has
a macroscopic $c \tau$. These points correspond to regions of the
parameter space in which $\Delta m$ is small, i.e., they are in the
co-annihilation regime.  Some of these points are within reach of
future direct detection experiments, which could lead to signals in
more than one kind of experiment. Other points, however, escape the
sensitivity even of ``ultimate'' detectors, such as
DARWIN~\cite{Schumann:2015cpa,Bottaro:2021snn}. This part of the
parameter space can still be probed at colliders, especially by taking
advantage of the long lifetime of $t ^{++}$. Note, in particular, that
all the points that evade future detection by DARWIN feature a
long-lived $t ^{++}$, leading to $c \tau \gtrsim 10$ m. This is why we
focus on the long-lived particle signatures in the following section.

\section{Collider phenomenology}\label{sec:ColliderPhenomenology}

\begin{figure}[ht]
	\centering
	\includegraphics[width=0.49\textwidth]{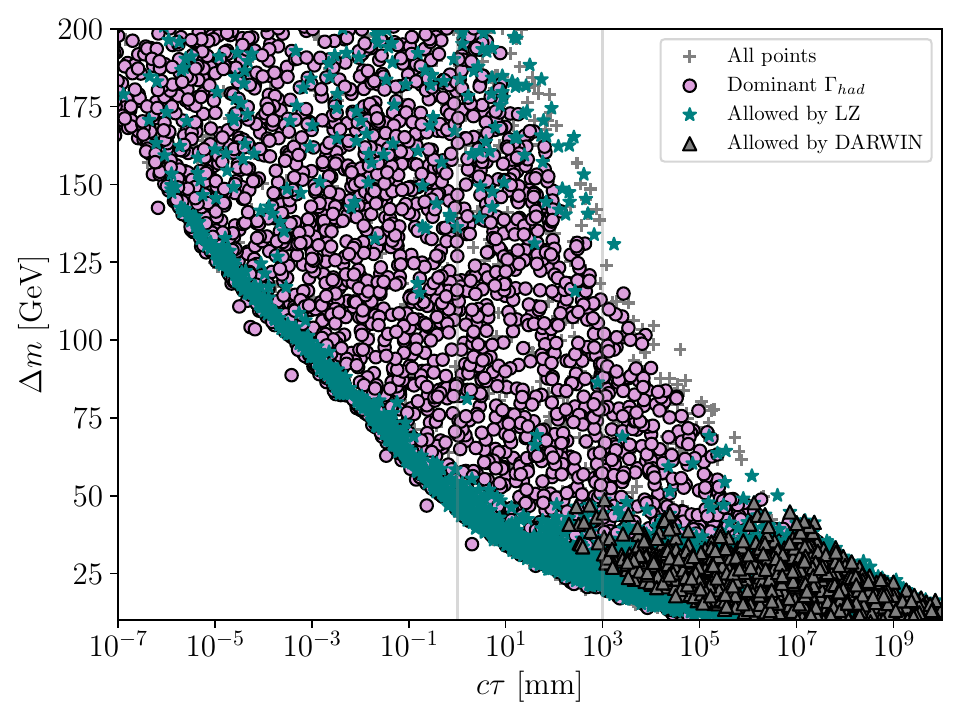}
	\includegraphics[width=0.49\textwidth]{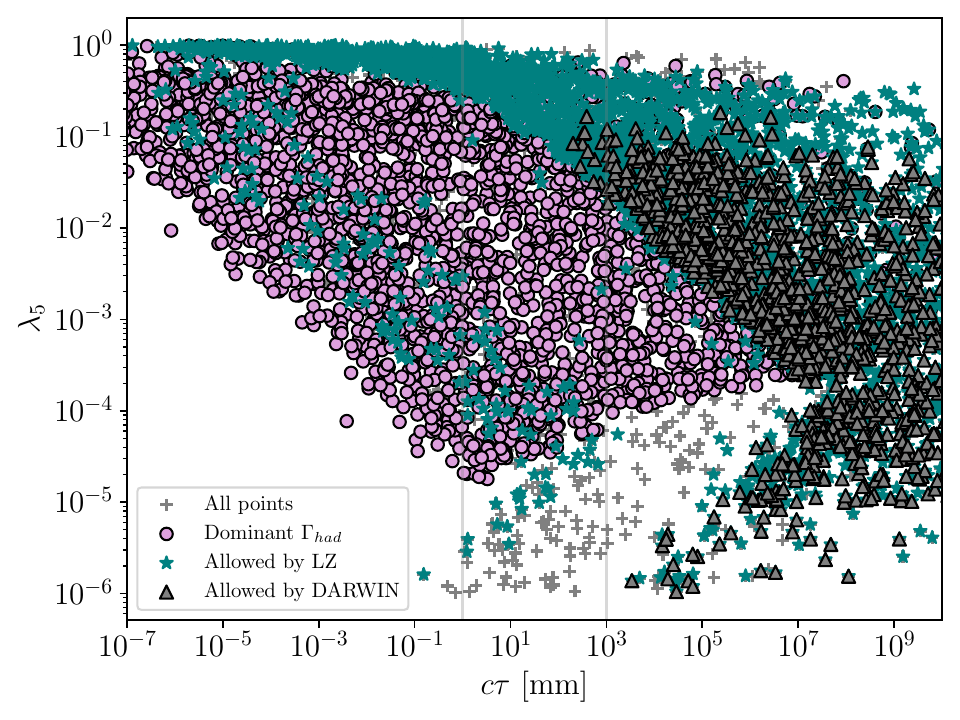}
	\caption{ Proper decay distance $c\tau$ as a function of $\Delta m$ (left panel), and $\lambda_{5}$ (right panel). All scanned points are represented in grey (with a cross) on the background. Points where the decays to hadronic $\Gamma_{had}$ dominates are shown in purple circles. We also overlay points satisfying the LUX-ZEPELIN (LZ) bound with a green star and DARWIN with a grey triangle. The 		``finite" $c\tau$ range is represented as vertical lines for $c\tau=1$ mm and $c\tau=1$ m.}
   \label{fig:DataCtauVsmt2}
\end{figure}

We study the collider signatures of the multi-charged candidate, $t^{++}$, which can have lifetimes ranging from the prompt regime to several hundreds of meters. Figure \ref{fig:DataCtauVsmt2} shows the lifetime dependence on the relevant model parameters for collider phenomenology \footnote{We fix for definiteness the mass of all heavy fermions to $m_{F}=2.5$ TeV, while $\lambda_{2},\lambda_{3a}=\lambda_{3b},\lambda_{6a}=\lambda_{6b}$ and $\lambda_7$ are free in what follows.}: $\lambda_5$ and $\Delta m\equiv m_{t^{++}}-m_{S_{0}}$, as a function of the proper decay distance of $t^{++}$, $c\tau$. Points where the decay width to hadrons ($\Gamma_{had}$) is dominant are shown in purple circles\footnote{For these points, the leptonic decay mode is suppressed, meaning, BR$(t^{++} \rightarrow l^{+} + l^{+} + S_{0}) < 0.1$}.

Macroscopic values of $c\tau$ starting from 1 mm and above -- for points that survive constraints from direct detection -- are achieved for values of $\lambda_{5}$ roughly smaller than $\sim 10^{-1}$ and $\Delta m \lesssim 50$ GeV. For larger values of $\Delta m$ points could still lie within the ``finite" $c\tau$ range (for $c\tau$ between 1 mm and 1 m, considering the acceptance of the ATLAS inner tracker detector), but then most of these points are excluded by direct detection experiments, as shown in the left panel of figure \ref{fig:DataCtauVsmt2}.

While some points in the prompt regime are still allowed by LZ, DARWIN would exclude them. This is why we do not consider collider constraints from prompt searches, as these points would already be excluded by future direct detection experiments (see also figure \ref{fig:DD_xsec_ms0}). We therefore focus on long-lived particle (LLP) searches because the reach is much bigger there than for the prompt case. 
In principle the LLP decay modes of $t^{++}$ include several 3-body and 5-body final states to DM, charged leptons, neutrinos and jets (see section~\ref{sec:Model}). We choose to focus on the delayed hadronic decay signatures of the $t^{++}$ ($\Gamma_{had}$, see figure~\ref{fig:crossingpoints}) at the LHC. Although we expect the reach in the leptonic channel (when considering $\Gamma_{ll}$) to be similar, we in fact expect a smaller reach with displaced leptons, when compared to displaced vertex searches in hadronic channels, where the vertex is made from charged particle tracks coming from quark hadronization~\cite{ATLAS:2015oan}. This can be expected due to larger backgrounds~\cite{CMS:2016isf,ATLAS:2020wjh} and/or softer displaced leptons~\cite{Blekman:2020hwr,ATLAS:2019fwx}, leading to smaller efficiencies~\footnote{Although we note in our model the leptons are not always soft, as $\Delta m$ could be large enough for the displaced lepton's transverse momentum to pass the trigger selections.}. We therefore choose to study the hadronic channel in detail.

Displaced vertex (DV) searches \cite{ATLAS:2017tny,CMS:2024trg} and searches for multi-charged particles \cite{ATLAS:2023zxo} will be sensitive to test the model at the LHC an its high-luminosity phase in the ``finite" $c\tau$ range i.e. $\sim \mathcal{O}(1 \text{mm} - 1 \text{m})$ and the ``larger" lifetime regime i.e. $c\tau \gtrsim \mathcal{O}(10 \text{m})$, respectively\footnote{Similar signatures for long-lived multi-charged particles predicted in neutrino mass models were studied in~\cite{R:2020odv,Hirsch:2021wge}.}. We consider the LHC production in {\texttt{MadGraph5\_aMC@NLOv3.5.1}}~\cite{Alwall:2011uj,Alwall:2014hca} of $t^{++}$ at $\sqrt{s}=13$ TeV, $pp\rightarrow t ^{++} t ^{--}$, with model spectra calculated from \texttt{Spheno}. We use the {\texttt{LUXqed17\_plus\_PDF4LHC15\_nnlo\_100}} parton distribution function set obtained from LHAPDF-6.2.3~\cite{Buckley:2014ana} to accurately account for photon-fusion diagrams. Generated events are then interfaced to \texttt{Pythia8}~\cite{Sjostrand:2014zea} for showering and hadronization. Missing transverse momenta, jets (with the help of \texttt{FastJet3.3.2}~\cite{Cacciari:2011ma}) and displaced vertices are also reconstructed within \texttt{Pythia8}, following a custom made code with displaced efficiencies validated for the ATLAS DV search previously in refs~\cite{Cottin:2018kmq,Proceedings:2018het}.

We start by reinterpreting two ongoing searches~\footnote{In principle, searches for disappearing charged tracks~\cite{ATLAS:2022rme,CMS:2023mny} could also constrain these types of models (see for instance an example in the Scotogenic model in ~\cite{Avila:2021mwg,Alguero:2021dig} ). Nevertheless, our phenomenology is more complex, as with a doubly charged LLP, the mass difference between the LLP and DM is large enough for our final state jets to be fully visible. We in addition have 5-body decays. We have checked for example constraints to our model from LLP searches contained within SmodelS~\cite{Alguero:2021dig}, and find it is not covered by any of the simplified model topologies. We thank Andre Lessa for his help in identifying this.} targeting the finite and large $c\tau$ regions: i) The ATLAS 13 TeV  ``DV+MET" search~\cite{ATLAS:2017tny} and the latest ii) ATLAS search for multi-charged particles~\cite{ATLAS:2023zxo}.

\begin{itemize}
\item i) The ATLAS 13 TeV ``DV+MET" search~\cite{ATLAS:2017tny} was first validated in \cite{Proceedings:2018het} by some of us. The signature includes at least one displaced vertex in association with large missing transverse momenta, used as a trigger. Following the recast prescription in the auxiliary material of the search \cite{atlas_DV_eff}, our event selection starts with the reconstruction of missing transverse momenta at the truth level, $p^{\text{miss}}_{T}>200$ GeV, which considers all stable and neutral BSM particles and SM neutrinos. We then require either one or two trackless jets with $p_{T}>70$ GeV or $p_{T}>25$ GeV, respectively.  A trackless
jet is a jet with a scalar sum of the $p_{T}$ of all of its charged particles inside it  to be less than 5 GeV. Following the prescription, these jet requirements are applied to 75\% of the events. We then apply cuts imposed on events with displaced vertices reconstructed from displaced tracks: displaced vertex coordinates within $4$ mm $< r_{\text{DV}} < 300$ mm and $|z_{\text{DV}}| < 300$ mm, number of final state charged tracks coming from the vertex to be at least 5 (each with $p_{T} > 1$ GeV and $|d_{0}| > 2$ mm), and invariant mass larger than 10 GeV (reconstructed assuming all tracks have the pion mass). We assume zero background, ensured by the large track multiplicity and invariant mass requirements of the displaced vertex.
In some of our search proposals in what follows, we tune the trigger cut and require a lower threshold of $p^{\text{miss}}_{T}>50$ GeV, while the displaced activity cuts remain the same as the original search.

\item ii) ATLAS also searched for multi-charged particles in ref.~\cite{ATLAS:2023zxo}. The search is also essentially background free for electric charges of $z\geq2$ and $c\tau \gtrsim 10 $ m. For the recast, we compare our cross-section results with the upper-limits presented by ATLAS in fig. 8 of ref.~\cite{ATLAS:2023zxo}, which shows the production cross-section as a function of mass for doubly charged particles. We estimate a lower limit on $t^{++}$ to be $m_{t^{++}}\gtrsim 800$ GeV for $\mathcal{L}=136$ fb$^{-1}$. As this search is essentially background free, we can estimate the future sensitivity by just scaling the existing limit with the expected luminosity gain. With this we get a lower limit of  $m_{t^{++}}\gtrsim 1240$ GeV at $\mathcal{L}=3000$ fb$^{-1}$. 
\end{itemize}

\begin{figure}[ht!]
	\centering
	\includegraphics[width=0.49\textwidth]{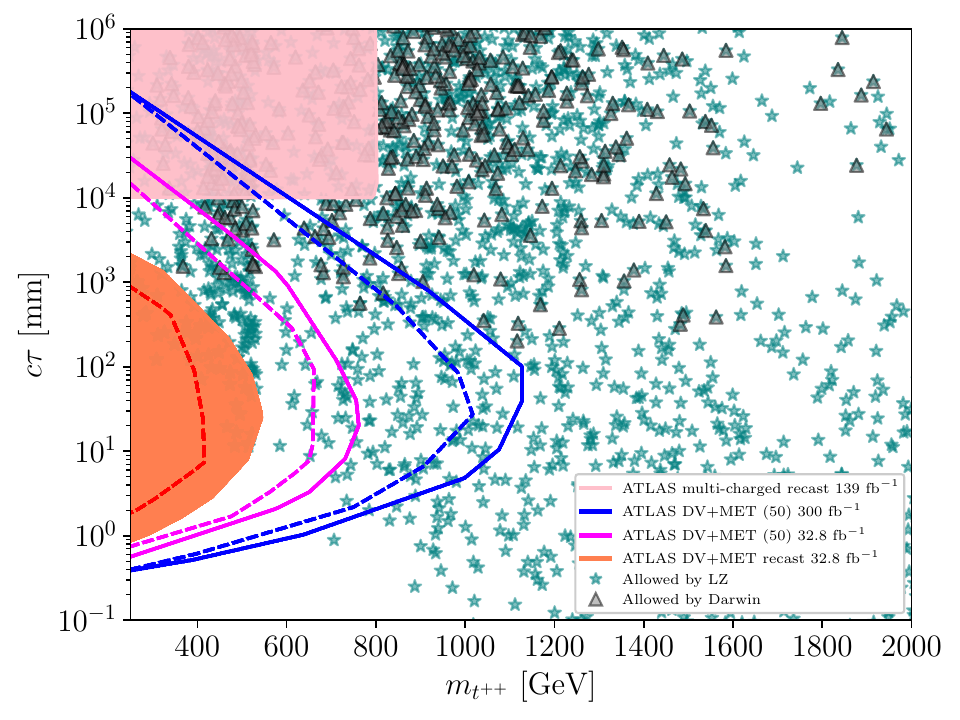}
	\includegraphics[width=0.49\textwidth]{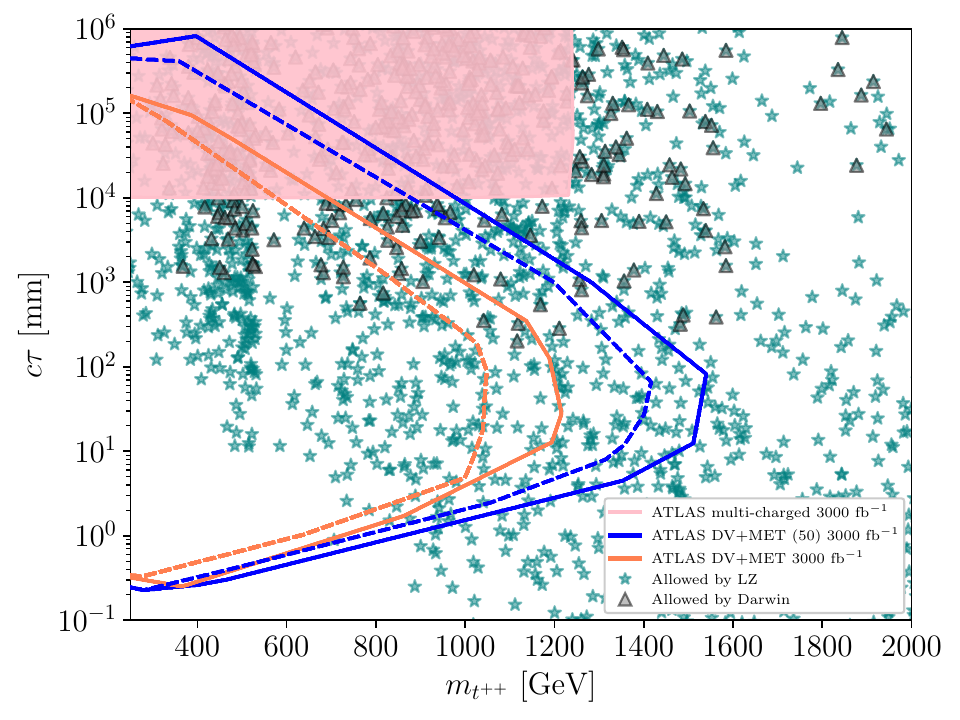}
	\caption{Projected LHC limits in the $m_{t^{++}}$ vs. $c\tau$ plane for long-lived particle searches with displaced vertices (solid and dashed lines), and multi-charged particles (pink shaded regions) for ``current" luminosities (left panel) and the higher-luminosity phase of the LHC (right panel). Solid lines assumes $100\%$ branching ratio of $\Gamma_{had}/\Gamma$ and dashed lines a $50\%$. Points satisfying the LUX-ZEPELIN (LZ) bound are marked with a green star and the ones satisfying DARWIN with a grey triangle.}
    \label{fig:LimitsCtauVsmt2}
\end{figure}

Figure~\ref{fig:LimitsCtauVsmt2} shows our ``DV+MET" reinterpreted limits in the plane $c\tau$ vs. $m_{t^{++}}$, by requiring 3 signal events (which correspond to 95 \% C.L. exclusion limits under zero background). The left panel shows in orange our straightforward recast at $\mathcal{L}=32.8$ fb$^{-1}$. We also extrapolate our limits to $\mathcal{L}=300$ fb$^{-1}$ with blue lines, considering a  proposal for a lower trigger threshold on missing transverse momenta of 50 GeV (instead of the original trigger of $p^{\text{miss}}_{T}>200$ GeV). In principle a tunning of the search cuts would require a proper experimental study of backgrounds, which is beyond the scope of this work. Nevertheless, our unique model signature consists of a doubly charged long-lived particle that can decay inside the trackers of LHC detectors~\footnote{Note that signal model benchmarks in searches for displaced vertices~\cite{ATLAS:2017tny,CMS:2024trg} usually target neutral long-lived particles, so our signature is not exactly covered by ongoing searches.}, and so the charged activity or passage through the detector can be further used to reconstruct additional observables such as large-ionization energy (similar as for multi-charged particle searches~\cite{ATLAS:2023zxo,CMS:2016kce} for instance) in order to eliminate possible background sources by enforcing additional displaced activity~\footnote{A similar direction for such an optimization in the context of Heavy Neutral Leptons was proposed in ref.~\cite{Cottin:2022nwp} by requiring a displaced trigger at CMS, which allowed for a lower $p^{\text{miss}}_{T}$ threshold.}. This motivates further experimental studies by the LHC collaborations particularly at the high-luminosity phase of the LHC. The right panel in figure~\ref{fig:LimitsCtauVsmt2} shows our DV projections for $\mathcal{L}=3000$ fb$^{-1}$ in orange lines, as well as the proposed search with $p^{\text{miss}}_{T}>50$ GeV. Masses of $t^{++}$ $\sim$ 1.5 TeV can be probed with this proposed optimization (about 300 GeV higher than what is achievable by projecting the limit of the original search). We also overlay in figure~\ref{fig:LimitsCtauVsmt2} derived limits from multi-charged searches in the pink region, which are sensitive to higher lifetimes for $c\tau> 10$ m and so complementary to searches for displaced vertices, which yield the strongest upper limit roughly at $c\tau\sim 10$ cm.

We also highlight that in figure~\ref{fig:LimitsCtauVsmt2}, solid lines assume a $100\%$ branching ratio to final states containing hadronic modes (i.e. $\Gamma_{had}/\Gamma$). This decay mode allows for plenty of final state visible tracks so that the multi-track displaced vertex search is applicable and efficient. We also present a conservative choice of $50\%$ branching ratio in dashed lines. 

From our results in figure~\ref{fig:LimitsCtauVsmt2} we draw that dark matter masses $m_{S_{0}}$ of $\mathcal{O}(1)$ TeV can be reached in the most optimistic scenario (blue solid line, left panel) at current LHC stages of the luminosity. This goes up to $\sim 1.4$ TeV at 3000 fb$^{-1}$ (blue solid line, right panel). Also, points that could be probed by multi-charged searches at the LHC could also be tested at future dark matter direct direction experiments as DARWIN. Points with dark matter masses above $\sim 1.4$ TeV and $\lambda_{5}\lesssim10^{-2}$ would still be allowed by DARWIN.

\section{Summary and discussion}\label{sec:conclusions}

In this paper we have studied the phenomenology of a neutrino mass
model, which we called SST for short, that explains at the same time
the dark matter problem. We have demonstrated that it is possible to
explain the relic density of DM and at the same time obey all existing
bounds from direct detection in this model. We note that the
combination of the different DM constraints favours a parameter
region, where the mass splitting between the scalar singlet and
triplet is much smaller than their average mass $\Delta m/m \ll 1$,
such that there is at least some contribution from co-annihilation
processes in fixing of the relic density. In this parameter region,
the doubly charged member of the triplet, $t^{++}$, typically has a
very small decay width.

The particular model variant considered in this paper, differs from
the famous \textit{scotogenic} model \cite{Tao:1996vb,Ma:2006km} in that
it has a much richer collider phenomenology. It is but one example in
a rather lengthy list of such one-loop neutrino mass DM models given
in \cite{Arbelaez:2022ejo}, albeit, as discussed below, most or all of
the collider signals we considered in this paper will be present in
many of the other models in this model class as well.  We stress
again, that one can consider the scotogenic model the proto-type in a
class of models that at the same time can explain neutrino masses and
dark matter, but \textit{all} models in this class can do so. The
motivation to study collider phenomenology arises from the fact that
the observation of a signal at the LHC (or any other future collider)
might be the only way to distinguish between all that plethora of
model variants.

Both, the neutral component of the fermion multiplet as well as the
lighter of the neutral scalars can be the DM, in principle.  In this
work, we have concentrated on the case of the scalar being the DM
(assuming that all fermions in the model are heavier than the scalar
states). The motivation for this choice is twofold. First, direct
detection constraints would rule out the neutral component of a
$F_{1,2,1/2}$ as DM unless a ($d=5$) non-renormalizable operator is
added to the model such that the fermion becomes "inelastic" dark
matter \cite{Bottaro:2022one}. To explain such a NRO, with the required 
size, would need an additional (particle) extension of the model.
Second, as discussed in the previous section, searches for long-lived
and/or quasi-stable particles are particularly promising in their mass
reach. While multiply charged scalars can have, in principle, any
decay length, for a charged fermion (member of a $SU(2)_L$ multiplet)
the situation is different. QED corrections split the masses of the
charged and the neutral fermions, such that the charged fermion will
always decay to the neutral fermion one plus a (very soft) charged
pion \cite{Cirelli:2005uq}. This gives an upper limit on the decay
length of charged fermions of at most a few cm for $SU(2)_L$ doublets
(and much smaller lengths for members of larger $SU(2)_L$ multiplets) 
\cite{R:2020odv}.

The phenomenology of multiply charged scalars depends on their decay
width. $t^{++}$ can decay either promptly, with a displaced vertex
(DV+MET search) or be quasi-stable (decay lengths larger than the
size of the detectors). Reinterpretation of existing DV+MET searches
\cite{ATLAS:2017tny, CMS:2024trg} leads to lower mass limits of order
${\cal O}(500-600)$ GeV, see fig. \ref{fig:LimitsCtauVsmt2}. We then
extrapolate these results to higher luminosities, projecting future
limits in the range $1.2-1.5$ TeV for 3000 fb$^{-1}$. A 
reinterpretation of ATLAS \cite{ATLAS:2023zxo} search for quasi-stable
multiply charged states allows us to estimate a lower limit on the
mass of $t^{++}$ to be around 800 GeV, based on a luminosity of 136
fb$^{-1}$. With ${\cal L}=3000 $fb$^{-1}$ this limit could be improved up to
$1.2$ TeV. 

As mentioned above, a comprehensive list of 1-loop neutrino mass
models that can also explain DM has been given in
\cite{Arbelaez:2022ejo}, encompassing a total of 318 model variants.
This large number prompts the question whether the conclusions drawn
in this article are applicable also to other models in this list.
Before closing, we would like to discuss this question at least
briefly, though only qualitatively.

First, we note that multiply charged scalars are a general feature
in this model class. From the 318 models, listed in the appendix
of \cite{Arbelaez:2022ejo}, 291 models contain multiplets with a
maximum charge of at least two. Even restricting ourselves only
to the more minimal set of models of table 8 of \cite{Arbelaez:2022ejo},
\footnote{Tables 9 and 10 contain models with larger representations
and hypercharges, which lead to even larger electric charges and production
cross sections, but admittedly look more contrived theoretically.}
we note that $S_{1,3,1}$ appears in 22 models, while 65 of the 91
models listed have doubly charged scalars, 18 of these models have
a triply charged scalar and there are even 2 models for which the largest
electric charge is actually 4.

The phenomenology of the doubly charged scalars will, in principle,
always be similar to what we discussed in this paper. In all cases the
decays of these states will contain either $WW$+MET or two same-sign
leptons plus MET. Again, depending on the mass splitting between
$S^{++}$\footnote{We use here $S^{++}$, to indicate any scalar with
  two electric charges, to differentiate from $t^{++}$, which is the
  doubly charged member of a triplet.} and the DM candidate, $S^0$,
the $W$-bosons could be on-shell or off-shell. In general, unless the
singly charged scalar (which must always be present too in models with
doubly charged states) is lighter than the $S^{++}$, one expects the
$S^{++}$ to be long-lived or quasi-stable, similar as in the SST
model. Production cross sections depend not only on the electric
charge but also on the $SU(2)_L$ multiplet, but for larger multiplets
than the triplet considered in this paper, larger production cross
sections are expected and, thus, the sensitivity reach of the LHC for
these should be correspondingly larger.

Cross sections at the LHC generically increase with electric charge.
For scalar states with similar masses one therefore always expects
that the state with the largest charge will be produced most
abundantly. Decay widths of scalars, on the other hand, become
drastically smaller with increasing electric charge. Here, two effects
are important. First, $n$-body phase spaces become smaller with
increasing $n$. And, second, in case of $n*W$ final states, the
available phase space, determined by $\delta m =m_{S^{n+}}-m_{S^0} -n
m_W$, will lead to a strong suppression of the width (for $\Delta m /m
\ll 1$\textit{}, as discussed above). For triply or even quadruply charged
states one can therefore expect that large parts of the parameter
space in these neutrino mass models, allowed by DM constraints, will
lead to quasi-stable scalars. The ATLAS search \cite{ATLAS:2023zxo}
shows {\em increasing} sensitivity to states with larger electric
charges, thus we can expect better sensitivities than what we found
in the SST model for models with highly charged scalars.

How can we distinguish between neutrino mass models in this list using
the LHC? Singling out {\em any} particular model will be hard.  Since
it is possible to determine the electric charge of long-lived scalars,
one could start to exclude models based on this rather straightforward
observation: The scotogenic model has no doubly charged particle, thus
the observation of a $S^{++}$ disfavours it. Observation of a triply
charged and a doubly charged scalar would rule out many more models
and so on.  However, if any of the signals studied in this article,
such as displaced vertices or multi-charged particles, were to be
observed, it would serve at least as a hint towards one of these
models. Additional information would be necessary to establish a
direct connection. For instance, in the case of signals with displaced
vertices where scalars decay within the detector, detecting different
decay modes ($WW +$ MET and two same-sign leptons + MET) would certainly
help towards identifying the underlying model.

This paper is mostly concerned with the determination of the parameter
space, where the $t^{++}$ predicted by our model could be discovered
at the LHC. Before closing, we would like to discuss briefly the
question, whether such a discovery could actually be used to
demonstrate that the neutrino mass generation mechanism has been
identified {\em experimentally}.

All Majorana neutrino mass models violate lepton number.  Thus, it
seems straightforward: Demonstrate that the particle discovered has
both, a lepton number conserving (LNC) and a lepton number violating
(LNV) decay mode. However, given the finite statistics expected even
for ${\cal L}=3/$ab, this requires that the hadronic and leptonic
final states of the $t^{++}$ decay have very similar branching ratios,
otherwise only one of the two will be observable. As
fig. (\ref{fig:crossingpoints}) shows, parameter points with
$\Gamma_{ll}=\Gamma_{had}$ exist. However, it is also obvious that
the ratio of these two widths can differ widely. Thus, the discovery
of lepton number violation is not guaranteed, even if the $t^{++}$ is
found at the LHC.

Even if both decay modes were measured at the LHC, there is the
problem that all final states contain a DM particle,
i.e. experimentally missing energy. LNV can not be easily established
in events with missing energy. To demonstrate that the missing energy
is not due to emission of neutrinos (which would make the leptonic
width actually a LNC mode) in the event is far from trivial too: One
would need to establish either that the escaping particle is massive
(SM neutrinos are ``massless'' for the LHC) or demonstrate (for
example, by studying the energy distributions of the charged leptons)
that only a single particle has escaped detection (and not two, as
would be necessary to compensate the lepton number of the two charged
leptons). In summary, it seems possible, in principle, to show
experimentally that the LHC has found LNV and thus traces of the
origin of neutrino mass. However, in practice to make all necessary
measurements requires that the model is realized in nature within some
particularly favourable part of parameter space.

To summarize, we have studied the dark matter and LHC phenomenology
of a neutrino mass model, which contains a doubly charged scalar.
Long-lived and quasi-stable particle searches at the LHC might
discover such a state in the future and thus, distinguish this
model from its famous scotogenic cousin. As we have briefly discussed
here, these LHC searches are actually sensitive to a large fraction
of all 1-loop neutrino mass models in this ``dark matter" class.

\bigskip

\section*{Acknowledgements}

G.C. thanks Gabriel Torrealba for his help with python and Andre Lessa
for useful comments. M.H. is supported by the Spanish grants
PID2020-113775GB-I00 (AEI/10.13039/ 501100011033), CIPROM/2021/054
(Generalitat Valenciana) and the MultiDark network, RED2022-134411-T.
G.C. acknowledges support from ANID FONDECYT grant No. 11220237.
J.C.H. acknowledges support from ANID FONDECYT grant No.1241685.
T.B.M acknowledges support from ANID FONDECYT grant No. 3220454.
G.C., J.C.H. and T.B.M. also acknowledge support from ANID -
Millennium Science Initiative Program ICN2019\_044. C.A. is supported
by ANID-Chile FONDECYT grant No. 1231248 and ANID-Chile PIA/APOYO
AFB230003.

\newpage


\providecommand{\href}[2]{#2}\begingroup\raggedright\endgroup

\end{document}

%% file: myTikz.tex
\usepackage{tikz}

\usetikzlibrary{decorations.markings}
\usetikzlibrary{decorations.pathmorphing}

\tikzset{
    vector/.style={decorate, decoration={snake}, draw},
    fermion/.style={draw=black, postaction={decorate},
        decoration={markings,mark=at position .55 with {\arrow[scale=1.,>=stealth]{>}}}},
    fermionbar/.style={draw=black, postaction={decorate},
        decoration={markings,mark=at position .58 with {\arrow[scale=1.,>=stealth]{<}}}},
    fermionline/.style={draw=black},
    gluon/.style={decorate, draw=black,
        decoration={coil,amplitude=4pt, segment length=5pt}},
    scalar/.style={dashed, draw=black, postaction={decorate},
        decoration={markings,mark=at position .55 with {\arrow[scale=1.,>=stealth]{>}}}},
    scalarbar/.style={dashed, draw=black, postaction={decorate},
        decoration={markings,mark=at position .58 with {\arrow[scale=1.,>=stealth]{<}}}},
    scalarline/.style={dashed,draw=black},
    fmassin/.style={draw=black, postaction={decorate},
    decoration={markings, mark=at position .75 with {\arrow[scale=1.,>=stealth]{>}}, mark=at position .30 with {\arrow[scale=1.,>=stealth]{<}}}},
}